\parskip=\normalbaselineskip
\parindent=0pt
 %------------------------------------------------------------------------
 \leftline{\bf Cosmology and Gravitostatics }
 \medskip
 \leftline{Peter Rastall}

 \leftline {Department of Physics and Astronomy, University of British
 Columbia, Vancouver BC, V6T 1Z1}

 \medskip
 ABSTRACT\par
  In gravitostatics, the minimally generalized Newtonian theory,
 the simplest cosmological model implies a cosmological redshift
 with acceleration parameter $-1$.  If the gravitational potential
 satisfies a wave equation, the total mass density (visible plus
 dark) is one third of the critical density, in good agreement
 with observation.  There is no horizon problem and no need to
 invoke dark energy.\par
 
{\it Subject headings:} gravitation --- cosmology: theory --- galaxies: high redshift --- dark matter --- early universe
\medskip

1. Introduction\par
\vskip-\parskip
The empirical evidence that might distinguish between different, relativistic theories of gravity is very sparse.  Any acceptable theory must of course be compatible with the Newtonian theory.  It used to be thought that the small deviations from the Newtonian predictions, the so-called 'classical tests of relativistic gravity', can be used to discriminate between relativistic theories.  However, a slightly modified  Newtonian theory, called {\it gravitostatics,} accounts for these deviations. Any relativistic theory must satisfy the classical tests, but this shows only that it is not incompatible with gravitostatics. (One must emphasize that gravitostatics is not a relativistic theory, although it reduces to special relativity in the limit when all gravitational fields vanish.)  To choose between different compatible theories, other experiments are needed.     \par   

In addition to the classical tests, there are other, small, 'postnewtonian'deviations from the Newtonian predictions, but they yield only weak constraints on possible theories. The most promising testing ground for relativistic theories is the study of gravitational radiation.  Although experiments are very difficult, there is one well-established, observational result: the rate of energy loss by gravitational radiation from a binary pulsar.  This agrees with the prediction of the Einstein theory, and is the only good piece of evidence for it at present.  The Einstein theory is widely accepted for aesthetic reasons, and because of the misapprehension that the classical tests support it uniquely, as discussed in the last paragraph.
\par
Cosmology is another field for the testing of gravitational theories.  The Einstein theory had difficulties here.  There was the 'horizon problem': spatially separated regions are causally disconnected at early times.  There was also the 'flatness problem': unless the initial conditions are chosen with extreme precision, the universe does not unfold as it should.  Both difficulties were ingeniously overcome by the theory of inflation.  It is still unclear whether this produces exactly the correct conditions for the subsequent expansion of the universe; even if it does, many feel that it is too contrived and that it destroys the aesthetic appeal of the theory.  Recently, measurements of very distant objects have shown that the Einstein theory is admissible only if the universe is filled with an otherwise unobservable 'dark energy'.  The invention of such entities is a mark of theories under strain: one thinks of phlogiston and the ether.
\par
In this uncertain situation, it is illuminating to see what gravitostatics says about cosmology.  This requires another small generalization of the theory.  the Poisson equation, inherited from the Newtonian theory, must be modified by the addition of a term involving time derivatives.  In this paper, the cosmological predictions of the modified theory are shown to be free from the problems of the conventional theory.  The way is now open for the investigation of properly relativistic theories with the same qualities.  
\par
The rationale for this paper is that, since the empirical evidence for the Einstein theory is not strong, it is worthwhile to investigate other theories.  This will reveal a wider range of possibilities and will help to distinguish real features of the world from merely theoretical constructs.  It is suggested that inflation and dark energy may be examples of such constructs.  The opposing view, more widely held, is that there is no point in developing other theories so long as observations do not clearly contradict the Einstein theory.  Which opinion one adopts is a matter of taste or temperament.  It requires cool assessment, not religious fervor.
\par
\medskip
2. Gravitostatics\par\vskip-\parskip
 It should be well known that the Newtonian theory of gravitation
 can be generalized to be compatible with special relativity in
 the limit of vanishing gravitational fields, and to account for
 the gravitational redshift and the other classical tests of
 relativistic gravity.  The generalized theory, {\it gravitostatics,} is non-speculative, in that it makes no new
 hypotheses but simply takes the old ones more seriously.  For
 example, one finds that the assumption that the gravitational
 potential is arbitrary to the extent of an additive constant
 implies that the rate of a clock depends exponentially on the
 gravitational potential --- and that the units of length and mass
 have a similar dependence.  A full account of all this can be
 found in (Rastall 1991); a shorter, more accessible source may be
 (Rastall 1975).  The next paragraphs try to summarize enough of
 the theory to make the rest of the paper comprehensible.
 \par
 
 In gravitostatics, there are preferred coordinate systems, called
 {\it Newtonian frames}, which are similar to the Galilean frames of
 Newtonian theory or the inertial charts of special relativity.
 As in Newtonian theory, the gravitational field is described by a
 function $\Phi $, the {\it gravitational potential.}  The measured values
 of universal  constants  such  as  the  speed of  light $c$ the
 Newtonian   gravitational  constant $G$, and
  Planck's  constant $h$ are assumed
 to   be  independent  of  $\Phi$. In a static  gravitational potential
 (that   is,  when $\Phi$ may  depend  on
  the  space  coordinates ${\bf x}=(x^{1},x^{2}, x^{3})$ of a Newtonian  frame  but  not
  on  the  time  coordinate $t$) the
 time   interval  measured  by a fixed  standard
  clock  is  not in  general   equal  to  the  change  in  the
  time  coordinate  $ t$.  Similarly,
 the   distance  as  measured  with a standard
  measuring  rod between
 the   spatial  points  with  coordinates ${\bf x}$ and ${\bf y}$ is  not in general  
$\vert{\bf  x}-{\bf y}\vert=[(x^{m}-y^{m})(x^m-y^{m})]^{1/2}$ (the
summation convention is assumed
unless stated otherwise: sum the repeated index $m$ over its range
$(1, 2, 3)$). The  spatial  geometry  as  measured by
standard  measuring rods is consequently    not  Euclidean. Directly
measured   times  and  lengths  will  be  termed
 {\it empirical};  those defined
in   terms  of  the  space  and  time
 coordinates  of a Newtonian chart will   be termed
{\it Newtonian}.\par

The energy of an object at any point can be measured in terms of
the energy of a reference object at a nearby point (the reference
object might be a standard slug of metal or a specified atom in
its ground state, for example).  This is called the {\it empirical
energy}.  From the laws of mechanics, which are a trivial
generalization of the special relativistic laws, it follows that
there is an energy function which is constant along the path of a
particle in a static gravitational potential.  This is called the
{\it Newtonian energy}, and is in general different from the empirical
energy.
\par
 
Empirical quantities may be denoted by a subscript $E$ and
Newtonian quantities by a subscript $N$.  For local quantities ---
those which are defined in a region where $\Phi $ is almost constant ---
one proves that the ratio of a Newtonian quantity to the
corresponding empirical quantity is an exponential function of $\Phi$
(this follows from the assumption that $\Phi $ is arbitrary to the
extent of an additive constant).  If $t_{N}$, $\ell_{N}$, $e_{N}$ are
a time
interval, length, and energy measured in Newtonian units, and $t_{E}$,
$\ell_{E}$, $e_{E}$ are the corresponding quantities in empirical units,
one has 
$$
t_{N}= t_{E}\exp[-\tau (\phi  - \phi _{0})],\qquad \ell _{N}= \ell
_{E}\exp[-\lambda (\phi  - \phi _{0})],
\eqno(1.1)$$
$$
e_{N}= e_{E}\exp[-\eta (\phi  - \phi _{0})],
$$
where $\tau$, $\lambda$, $\eta $, and $\phi _{0}$ are constants, 
$\phi  = \Phi /{c_E}^2$, and $c_{E}$ is the
speed of light in empirical units.  One calls $\phi $ the {\it
dimensionless potential.}  Newtonian and empirical values are equal when 
$\phi  = \phi_{0}$;
a Newtonian frame for which this is true is called 
a $\phi _{0}$ {\it frame}.
\par

The magnitude of a dimensionless quantity is independent of the
units of measurement.  It follows that, when measured in
Newtonian units, the magnitudes of two quantities with the same
dimensions must have the same dependence on $\phi$  (because their
ratio is dimensionless).  Since the dimensions of a mass $m$ are
$[m] = [ e \ell ^{-2}t^{2}]$, eq.(1.1) gives
$$m_{N}= m_{E}\exp[-\mu (\phi  - \phi _{0})],
\eqno(1.2)$$
where $\mu  = \eta  - 2\lambda  + 2\tau $.  
To avoid an excess of subscripts, one makes the convention that
universal constants without a subscript $E$ or $N$ are to be
interpreted as empirical quantities (the speed of light $c$ is to
be interpreted as $c_{E}$, Planck's constant $h$ as $h_{E}$, etc.).
Anything else written without a subscript $E$ or $N$ (i.e. anything
but a universal constant) is to be interpreted as a Newtonian
quantity, unless stated otherwise (the velocity ${\bf V}$ of a particle
is to be interpreted as ${\bf V}_{N}$, etc.).
\par
 
We have now set up the conceptual framework of gravitostatics.
To go further, we must learn how to calculate the motion of a
particle in a gravitational field.  This would seem to require
some additional, arbitrary hypothesis.  However, the Lagrangian
method allows us to proceed in a non-arbitrary, or at least very
natural way.
\par
 
The Lagrangian of a free particle in special relativity is 
$L = -mc^{2}/\gamma $, where $m$ is the constant and invariant proper mass,
$\gamma  = (1
- V^{2}c^{-2})^{-1/2}$, and $V = \vert {\bf V}\vert $ is the speed.  The
guiding principle of
gravitostatics is to adopt the laws of special relativity and
Newtonian gravity with minimal changes.  One therefore assumes
that the Lagrangian of a particle subject only to a gravitational
field is the same as that for a free particle --- except that
everything is to be measured in Newtonian units.  With the
previous conventions, one has
$$
L = - m{c_N}^2/\gamma ,\qquad \gamma  = (1 - V^2{c_N}^{-2})^{-1/2},
\eqno(1.3)$$
where $m$ is the proper mass in Newtonian units.  As in Newtonian
mechanics, it is assumed that $m_{E}$, the proper mass in empirical
units, is constant along the path. (It would of course be more
correct to write the Lagrange function as $L({\bf Z}, {\bf V}, \cdot)$, where
${\bf Z}$ is
the position of the particle, and its value at time $t$ as $L({\bf Z}(t)$,
${\bf V}(t),t)$.)  From (1.1), one gets 
$$ c_N= c \exp[(\tau  - \lambda )\psi ],\qquad 
m{c_N}^2 = m_{E}c^{2}\exp(-\eta \psi ),\eqno(1.4)$$
where $\psi  = \phi  - \phi _{0}$.  If $\psi $ and $V$ are small, the
Lagrangian must be
equivalent to the Lagrangian $m_{E}V^{2}/2 - m_{E}c^{2}\phi $ of Newtonian
gravitation, which implies that $\eta  = -1$.
\par

The $r$ component of the momentum of the particle is 
$p_{r}=\partial L/\partial V^{r}= m\gamma V^{r}$ and the Euler-Lagrange
equation is $dp_{r}/dt = \partial L/\partial Z^{r}$. 
Since $V^{r}p_{r}- L = m\gamma (V^{2}- {c_N}^2\gamma ^{-2}) = m\gamma
{c_N}^2$, one has
$$(d/dt)(m\gamma {c_N}^2) = -\partial _{t}L,\eqno(1.5)$$
where $\partial _{t}L$ denotes the partial derivative with respect to the
last
variable in $L = L({\bf Z}, {\bf V}, \cdot)$.  If $\phi $ is static, then
$\partial _{t}L = 0$, and $E
= m\gamma {c_N}^2$ is constant along the path of the particle.  It is
identified with the {\it energy} of the particle in Newtonian units.
The Newtonian unit of energy can thereby be defined everywhere in
the Newtonian frame.  One may choose the Newtonian unit to be the
same as the empirical unit in a region where the potential is $\phi _{0}$.
(If $\phi $ is quasistatic, as will later be allowed, then (1.5) gives
the change in $E$ along the path.)
\par
 
Rapidly moving objects are permissible in gravitostatics,
provided that they do not contribute significantly to the
gravitational field.  One can introduce a classical model of a
photon as the limiting case of a particle whose mass tends to
zero and whose speed tends to the speed of light in such a manner
that its energy $E = m\gamma {c_N}^2$ remains bounded.  The momentum of the
photon is parallel to its velocity and has magnitude $E/c_{N}$.  In a
static potential, the Newtonian energy $E$ of the photon is constant
along its path.
\par
A photon of Newtonian energy $E$ is associated with an
electromagnetic wave of Newtonian frequency $\nu $.  Planck's law
states that $E_{E}/\nu _{E}= h$, where $h$ is Planck's constant, or
equivalently that $E/\nu  = h_{N}$.  In a static potential, the Newtonian
energy $E$ of the photon is constant along its path, and the
Newtonian frequency $\nu $ of the associated wave is constant
   $_{}($otherwise,  the  number of wavelengths
 between  two spatial
points is not constant in time).  Hence $E/\nu $ is constant along the
path.
At the potential $\phi _{0}$, where Newtonian units are the same as
empirical, one has $E/\nu  = h$ and it follows that, at any potential,
$E/\nu  = h_{N}= h$: the value of Planck's constant is the same in
Newtonian and empirical units.  Since $h_{N}= h \exp[-(\eta  + \tau )\psi
]$, from
(1.1), we have $\tau  = -\eta  = 1$.
\par

In the Newtonian theory, the gravitational potential $\Phi $ obeys
Poisson's equation $\partial _{m}\partial _{m}\Phi  = 4\pi G\rho $, where
$G$ is the gravitational
constant and $\rho $ is the mass density of the sources.  In
gravitostatics, the only difference is that one uses the energy
density $\epsilon $ of the sources:
$$
\partial _{m}\partial _{m}\phi  = 4\pi k_{N}\epsilon ,
\eqno(1.6)$$
where $\phi  = \Phi c^{-2}$ and $k_{N}= G_{N} {c_N}^{-4}$, or $k = Gc^{-4}($note
that $\Phi $ is
measured in empirical units).
\par
 
In Newtonian theory, the active gravitational mass of an object,
which determines the strength of the gravitational field which it
produces at large distances, is the same as the passive
gravitational mass (the weight) and the inertial mass.  In
gravitostatics, similarly, the active gravitational energy is
assumed to be equal to the total energy of the sources.  A short
calculation shows that this implies $\lambda  = \eta $.  All three
constants in (1.1) are now determined:
$$\tau  = -\lambda  = -\eta  = 1,\eqno(1.7)$$
and eqs.(1.4) and (1.3) become 
$$c_{N}= ce^{2\psi }, \qquad m{c_N}^2= m_{E}c^{2}e^{\psi },\qquad L =
-m_{E}c^{2}e^{\psi }(1 - V^{2}c^{-2}e^{-4\psi })^{1/2}. \eqno(1.8)$$
Using the values (1.7), one calculates the gravitational
redshift, the bending of light by the Sun (and the
related radar time-delay), and the anomalous perihelion shift of
Mercury.  These are often called the {\it classical tests of
relativistic gravity}.  In fact, they are predicted by
gravitostatics, which is only a slight generalization of the
Newtonian theory.
\par

\medskip
3. The cosmological redshift\par\vskip-\parskip
Like  Newtonian  theory,  gravitostatics  can
 be  applied when the   gravitational  field  is  slowly  varying
 in   time. The simplest   cosmological  models  are  those  in
 which the gravitational   potential  is  independent  of  the
 space  coordinates  and is
a function only of   the  time  coordinate $t$ of a Newtonian  frame.
We   take  the  special  case  when  the
 potential  is a linear function of   the  empirical time $T$:
$$\psi  = \phi  - \phi _{0}= -HT,\eqno(2.1)$$
where $\phi _{0}$ and $H > 0$ are constants and $T$ is a function of
$t$ only.
Since $T$ at any spatial point can be regarded as time measured by
a standard clock fixed at that point, one has $dT/dt = e^{\psi }= e^{-HT}$
 from (1.1) with $\tau  = 1$.  Integrating and choosing $T = 0$ when $t =
0$, one gets $e^{-\psi }= Ht + 1$.  As $t \longrightarrow  \infty ,
\psi  \longrightarrow  -\infty $ and $T \longrightarrow  \infty $;
as $t \longrightarrow
-1/H, \psi  \longrightarrow  \infty $ and $T \longrightarrow
-\infty $.  In this model the `horizon problem' of
cosmology does not arise: since $c_{N}= ce^{2\psi }$, signals can, in
principle, be sent to any spatial point ${\bf x}$ at a time $t$ from any
other spatial point ${\bf x}^\prime $ at a time $t^\prime $, provided that
$t^\prime $ is
sufficiently close to the value $-1/$H.
\par

To define the redshift, we suppose that light emitted by an atom
at rest at the  spatial point 1 in a Newtonian chart has frequency $\nu _{E}$,
measured in empirical units.  It travels to the spatial point 2,
where its frequency is measured to be $\nu _{2E}$.  This is compared with
the frequency $\nu _{E}$ of the light emitted from an identical atom at
rest at the spatial point 2.  The redshift $z$ (of the light from 1
as measured at 2) is defined by
$$z = (\nu _{E}- \nu _{2E})/\nu _{2E}= (E_{E}-
E_{2E})/E_{2E},\eqno(2.2)$$
where the frequency $\nu _{E}$ of the light and the energy $E_{E}$ of the
corresponding photon are related by $\nu _{E}= E_{E}/h$, and similarly $\nu
_{2E}= E_{2E}/h$.
\par
 
It is easiest to calculate the redshift by again considering the
motion of a classical `photon'.  The Newtonian energy of such a
photon changes according to (1.5).  From (1.8) one has $\partial
L/\partial \psi  =
-E(1 + V^{2}{c_N}^{-2})$, and putting $V = c_{N}$ gives
$$dE/dt = 2E\partial _{t}\psi .\eqno(2.3)$$
As before, a photon of empirical energy $E_{E}$ is emitted by an atom
at rest at the spatial point 1 at the instant $t_{1}$ in a Newtonian
chart.  The Newtonian energy of the photon is $E_{1}= E_{E}e^{\psi _{1}}$,
from
eqs.(1.1) and (1.7), where $\psi _{1}= \phi _{1}- \phi _{0}$ and $\phi
_{0}$ is a constant.
The photon travels to the point 2, where its Newtonian energy is
$E_{2}$.  From (3.3), $\ln(E_{2}/E_{1}) = \int 2 \partial _{t}\psi ({\bf
Z}(t), t)dt$, where the integral
is from $t_{1}$to $t_{2}$.  The photon's empirical energy at point 2 is
\par
$$E_{2E}= E_{2}e^{-\psi _{2}}= E_{E}\exp[\psi _{1}- \psi _{2}+ \int^{t{ }
_{2}}_{t{ } _{1}}2\partial _{t}\psi ({\bf Z}(t), t)dt].
\eqno(2.4)$$
 From (3.2), the redshift is 
$$z = (E_{E}/E_{2E}) - 1 = \exp[\psi _{2}- \psi _{1}- \int^{t{ }
_{2}}_{t{ } _{1}}2\partial _{t}\psi ({\bf Z}(t), t)dt] - 1. \eqno(2.5)$$
In the special case of a static potential, one has $\partial _{t}\psi  =
0$ and
the redshift is $z = \exp(\psi _{2}- \psi _{1}) - 1 \approx  \psi _{2}-
\psi _{1}$ if $\vert \psi _{2}- \psi _{1}\vert  \ll  1$.
Note that the redshift is positive if the light goes from a lower
to a higher potential.  If $\partial _{m}\psi  = 0$ everywhere, so that
$\psi $ is a
function of $t$ only, eq.(2.5) becomes
$$
z = \exp(\psi _{1}- \psi _{2}) - 1.
\eqno(2.6)$$
Here the redshift is positive if the light goes from a higher to
a lower potential.
\par
 
In the special case when $\psi  = -HT$, eq(2.1), the redshift is
$$
z = \exp[H(T_{2}- T_{1})] - 1 = H(T_{2}- T_{1}) + (1/2)H^{2}(T_{2}-
T_{1})^{2}.
\eqno(2.7)$$
The last equation is valid if $H(T_{2}- T_{1})$ is small.  The term in
$H^{2}$ corresponds to a deceleration parameter $q = -1$, in the
conventional terminology (see eq. (6.2) or eq.(29.15) of (Misner,
Thorne, \& Wheeler 1973)). Eq.(2.7) implies that the universe is
`accelerating' --- it expands more rapidly at later times.  This
agrees with recent measurements on the redshifts of distant
supernovae and galaxies (Knop, et al. 2003), (Tonry, et al.
2003), (Daly \& Djorgovski 2003).
\par
\medskip
4. A field equation and dark mass

The field equation (1.6) is the same as the classical Poisson
equation, except that the mass density of the sources is replaced
by their energy density.  An apparent defect of the equation is
that the energy density of the gravitational field does not
appear in the source term.  It is however easy to remedy this:
instead of choosing the field variable to be $\phi $, or $\psi  = \phi  -
\phi _{0}$,
one takes it to be $e^{\psi /2}$.  One then has $2e^{-\psi /2}\partial
_{m}\partial _{m}e^{\psi /2}= \partial _{m}\partial _{m}\psi  +
(1/2)\partial _{m}\psi \partial _{m}\psi $ and (1.6) becomes
$$
2e^{-\psi /2}\partial _{m}\partial _{m}e^{\psi /2}= 4\pi k_{N}(\epsilon
+ \epsilon _{G}),
\eqno (4.1)$$
where $\epsilon _{G}= (8\pi k)^{-1}\partial _{m}\psi \partial _{m}\psi
$ is the energy density of the
gravitational field.  Since the dimensions of $k = Gc^{-4}$ are length
divided by energy, we have $k_{N}= k$, from (1.1) with $\eta  = \lambda  =
-1$.
Note that $\epsilon _{G}$ is never negative --- unlike the energy density in
Newtonian gravitation, which is $-\epsilon _{G}$.
\par
 
Eq. (4.1) is completely equivalent to (1.6).  It is the equation
satisfied by a static potential $\psi $.  If the potential is not
static, then the field equation must involve time derivatives if
it is to determine the time evolution of the potential.  Here a
speculative element enters the theory.  We guess, as the most
obvious possibility, an equation similar to the wave equation:
$$
2e^{-\psi /2}[\partial _{m}\partial _{m}e^{\psi /2}- {c_N}^{-1}\partial
_{t}({c_N}^{-1}\partial _{t}e^{\psi /2})] = 4\pi k(\epsilon  + \epsilon _{G}).
\eqno (4.2)$$
(The $c_{N}$ are inserted to ensure that the equation is invariant
under change of $\phi _{0}$ chart.)
\par

If we assume that $\psi (t) = - HT$, as in (3.1), and use $c_{N}=
ce^{2\psi }$ and
$dT/dt = e^{\psi }$, we have 
$$2e^{-\psi /2}{c_N}^{-1}\partial
_{t}({c_N}^{-1}\partial _{t}e^{\psi /2}) = -(H^{2}/2c^{2})e^{-2\psi }.$$
Defining the {\it critical density} $\epsilon _{c}$ by $\epsilon _{cE}=
3H^{2}/8\pi kc^{2}$, and $\epsilon _{c}=
\epsilon _{cE}e^{-2\psi }$, we find that  (4.2) becomes
$$
2e^{-\psi /2}\partial _{m}\partial _{m}e^{\psi /2}= 4\pi k(\epsilon
+ \epsilon _{G}- \epsilon _{c}/3).
\eqno (4.3)$$
Since the space derivatives vanish in this cosmological model, we
have $\epsilon _{G}= 0$ and $\epsilon  = \epsilon _{c}/3$.
\par

The average density of the visible matter in the universe is only
a few percent of the critical mass density.  Observations of the
motions of galaxies and of gravitational lensing effects, etc.,
show that `dark' gravitating matter must be present.  The average
total density of visible plus dark matter is about 30 percent of
the critical mass density.  If we identify $\epsilon  = \epsilon
_{c}/3$ as the
average energy density of all the matter, visible plus dark, we
have quite good agreement with observation.  The theory does not say
anything about the nature of the dark mass.
\par
\medskip
5. Conclusions

As emphasized earlier, gravitostatics is not a new, speculative
theory, but only a minimally modified form of Newtonian
gravitation.  In this paper, we have generalized it slightly by
allowing gravitational fields to vary slowly in time, and have
assumed that they satisfy a wave equation.  The simplest solution
of this equation implies that the average total mass density is
one-third of the critical mass density, in  agreement
with observation.  The universe has an infinite past, as measured
in empirical units, and there is no horizon problem.  The
deceleration parameter has the value $-1$, which implies that the
expansion rate of the universe increases with time, and which
accords with observations of objects with high redshift.  One
does not have to postulate a mysterious `dark energy'.
\par
 
An obvious generalization of the results is given in the
following Appendix.  We showed above that if $\psi  = -HT$ then the
time-dependent term in the field equation  (4.2) gives rise to a
contribution to the energy density of the sources which is
constant in time when measured in empirical units.  If we now
write the time-dependent term as $2e^{-\psi /2}{c_N}^{-1}\partial
_{t}({c_N}^{-1}\partial _{t}e^{\psi /2}) =
c^{-2}e^{-2\psi }B$, our result is that $B$ is constant if $\psi  = -HT$.
One
naturally asks what is the most general $\psi $ that implies a constant
$B$.  We find that, in the general case, the contribution to the
energy density may be less than before.  The deceleration
parameter may depend on the time of observation --- it lies between
$-1/2$ and $-1$ and tends to the latter value at large times.
\par
 
This paper uses the conventional model of a universe with
homogeneous space sections, which cannot be expected to accord
exactly with obervation.  The equations of any such model will
differ from those found by averaging over our actual, lumpy
universe, because the non-linear terms will almost certainly not
average to zero.  Once the distribution of dark matter is known,
it should be possible to estimate the size of the error.
\par
 
In summary, we have shown that a simple, rather primitive theory
of gravitation can account for the large-scale structure of the
universe without invoking hypotheses such as inflation or dark
energy.  The challenge now, for vigorous young theorists, is to
find a properly relativistic theory that is at least as
successful. This is not trivial because one cannot assume that the potential is a relativistic invariant (it is well known that relativistic scalar theories of gravity do not work).  The examples of exact theories in Chapter IX of (Rastall 1991) do not take account of the field equation (4.2).
\par
\medskip
6. Appendix

We shall find the most general function $\psi $ that depends only on
the time coordinate $t$ and that satisfies the equation
$2e^{-\psi /2}\partial_t (e^{-3\psi /2}\partial_{t}\psi ) = B$,
where $B$ is a constant. We write $w =
e^{-3\psi /2}$ and the equation becomes $d^{2}w/dt^{2}= -(3B/2)w^{-1/3}$.
Multiplying by $dw/dt$ and integrating gives $dw/dt =
\pm (a + bw^{2/3})^{1/2}$, where $b = -9B/2$ and $a$ is a constant of
integration.  The upper sign is chosen so that $d\psi /dt < 0$.
Defining $u = w^{-2/3}= e^{\psi }$, gives $t = -(3/2) \int u^{-2}(au
+ b)^{-1/2}du$,
which is a standard integral.
\par

The proper time $T$ satisfies $dT/dt = e^{\psi }$.  Since $dt/du =
-3/2u^{2}S$,
where $S = (ae^{\psi }+ b)^{1/2}$, and $du/d\psi  = u$, we have $dT/d\psi
= -3/2S$, and
hence $\psi ^\prime = d\psi /dT = -2S/3$, 
$\psi ^{\prime\prime} =(2/9)ae^{\psi }$. Eq.(2.6) implies that
the redshift observed at time $T_{2}$ is
$$
z = -\psi ^\prime (T_{2}- T_{1}) + (1/2)(\psi ^{\prime ^{2}}+ \psi
^{\prime\prime})(T_{2}- T_{1})^{2},
\eqno (6.1)$$
provided that $T_{2}- T_{1}$ is small, where $\psi ^\prime $ and $\psi
^{\prime\prime}$ are to be
evaluated at $T_{2}$.  The conventional form of this redshift is
$$z = H(T_{2}- T_{1}) + (1 + q/2)H^{2}(T_{2}- T_{1})^{2},
\eqno (6.2)$$
where $H$ and $q$ are the Hubble parameter and deceleration parameter
at $T_{2}$.  It follows that $H = 2S/3$ and $q = -1/2 - b/2S^{2}$.  Note
that
$q$ lies between $-1/2$ and $-1$, and that $q$ tends to $-1$ when $T_{2}$ is
large.  If $T_{2}= T_{0}$, the present epoch, and if the Newtonian frame
is chosen so that $\psi (T_{0}) = 0$, then Newtonian and empirical units
coincide at $T_{0}$, and $S(T_{0}) = (a + b)^{1/2}$.  Writing $H =
H_{0}$ and $q = q_{0}$, we have
$$a = (9/2)H_{0}(1 + q_{0}), \qquad b = -(9/4)(1 + 2q_{0})H_{0}.\eqno (6.3)$$
This determines $a$ and $b$ in terms of the measurable $H_{0}$ and
$q_{0}$.  
the earlier assumption that $\psi  = -HT$, eq.(2.1), corresponds to $q_{0}
= -1$.
\par
\medskip
REFERENCES

\parskip=0pt
Daly, R. A. \& Djorgovski, S. G. 2003, ApJ, 597, 9\par
Knop, R. A., et al. 2003, ApJ, $598, 102$\par
Misner, C. W., Thorne, K. S., \& Wheeler, J. A. 1973, Gravitation
(San Francisco: W. H. Freeman)\par
Rastall, P. 1975, Am. J. Phys., $43, 591$\par
Rastall, P. 1991, Postprincipia: Gravitation for Physicists and
Astronomers (Singapore: World Scientific)\par
Tonry, J. L., et al. 2003, ApJ, 594, 1
\bye